\documentclass[sigconf]{acmart}

\usepackage{algorithmic}
\usepackage[algo2e,ruled,vlined]{algorithm2e}
\SetCommentSty{scriptsize}
\SetKwComment{tcc}{\{}{\}}
\SetKwIF{If}{ElseIf}{Else}{if}{}{else if}{else}{endif}
\SetKwFor{While}{while}{}{endw}
\SetKwFor{ForEach}{for each}{}{endfch}
\newtheorem{exmp}{Example}[section]
\SetKwRepeat{Do}{do}{while}

\usepackage{todonotes}
\usepackage{subcaption}

\title{Optimizing Irregular Communication with Neighborhood Collectives and Locality-Aware Parallelism}

\author{Gerald Collom}
\affiliation{
    Department of Computer Science
    \institution{University of New Mexico}
    \city{Albuquerque}
    \country{USA}
}
\author{Rui Peng Li}
\affiliation{
    Center for Advanced Scientific Computing
    \institution{Lawrence Livermore National Laboratory}
    \city{Livermore}
    \country{USA}
}
\author{Amanda Bienz}
\affiliation{
    Department of Computer Science
    \institution{University of New Mexico}
    \city{Albuquerque}
    \country{USA}
}

\begin{document}

\begin{abstract}
Irregular communication often limits both the performance and scalability of parallel applications.  
Typically, applications individually implement irregular messages using point-to-point communications, and any optimizations are added directly into the application.  
As a result, these optimizations lack portability.  
There is no easy way to optimize point-to-point messages within MPI, as the interface for single messages provides no information on the collection of all communication to be performed.  
However, the persistent neighbor collective API, released in the MPI 4 standard, provides an interface for portable optimizations of irregular communication within MPI libraries.

This paper presents methods for optimizing irregular communication within neighborhood collectives, analyzes the impact of replacing point-to-point communication in existing codebases such as Hypre BoomerAMG with neighborhood collectives, and finally shows an up to 1.32x speedup on sparse matrix-vector multiplication within a BoomerAMG solve through the use of our optimized neighbor collectives.  
The authors analyze multiple implementations of neighborhood collectives, including a standard implementation, which simply wraps standard point-to-point communication, as well as multiple implementations of locality-aware aggregation.  
All optimizations are available in an open-source codebase, MPI Advance, which sits on top of MPI, allowing for optimizations to be added into existing codebases regardless of the system MPI install.
\end{abstract}

\maketitle


\section{Introduction}
Parallel applications, such as simulations and iterative solvers, are often bottlenecked by irregular point-to-point communication. 
For instance, the performance and scalability of Hypre~\cite{hypre, BoomerAMG}, a widely-used algebraic multigrid (AMG) solver, is limited by the irregular communication that occurs throughout the numerous sparse matrix operations.  
While there are many optimizations for point-to-point communication, including persistent communication and locality-aware aggregation, there is no widely used library supporting these optimizations, requiring each application to optimize code by hand.  
This paper presents optimizations of point-to-point communication within MPI neighborhood collectives and analyzes the performance of these operations within the Hypre BoomerAMG solver.

Each parallel application typically implements their own irregular communication with calls to \texttt{MPI\_Isend} and \texttt{MPI\_Irecv}, or some variation of these methods.  
Communication optimizations are currently added within applications, and as a result not easily shared among parallel codebases.  
For instance, the AMG solvers Hypre, Muelu~\cite{MueLu}, and GAMG~\cite{petsc} each call separate implementations for point-to-point communication within sparse matrix operations, with optimizations unique to each.  
Furthermore, there is no easy way to add point-to-point optimizations within methods such as \texttt{MPI\_Isend} and \texttt{MPI\_Irecv} as these only pass information about a single message rather than the collection of all messages.  

This paper addresses the point-to-point communication bottleneck through the use of MPI neighborhood collectives, which wrap irregular communication and allow for optimizations within MPI.  
The sparse collective interface requires applications to provide information about all messages, allowing for optimizations within the method.  
While neighborhood collectives provide sufficient information for optimizations within MPI, many communication optimizations incur large initial overheads which are offset during subsequent iterations.  
Therefore, the addition of persistent neighborhood collectives in the MPI 4 standard allows for substantial irregular communication optimizations to be added within MPI.  
Adding these optimizations within MPI implementations will allow for all applications to take advantage of them by simply calling the appropriate neighborhood collective.

While neighborhood collectives have potential to alleviate critical communication bottlenecks in irregular applications, they have yet to be widely adopted. 
While the interface has existed since the MPI 3 standard, implementations of neighborhood collective often simply wrap point-to-point communication with limited exploration of possible optimizations.  
As few applications use these methods, there is little incentive to improve optimizations.  
At the same time, while the implementations contain few optimizations, there is little advantage to rewriting existing applications to utilize these methods.  
The goal of the work presented in this paper is two-fold: to create an optimized implementation of the persistent version of the neighborhood collective \texttt{MPI\_Neighbor\_alltoallv}, and to restructure existing parallel codebases, such as the widely used parallel multigrid solver Hypre, to replace point-to-point communication with persistent neighborhood collectives.  
All neighborhood implementations are added to a lightweight library, called MPI Advance~\footnote{https://github.com/mpi-advance}, which sits on top of MPI, allowing it to optimize which system version of MPI is installed.  
Furthermore, all neighborhood collective additions to BoomerAMG are published in the neighbor collective branches of Hypre.

\begin{figure}[ht!]
    \centering
    \includegraphics[width=\linewidth]{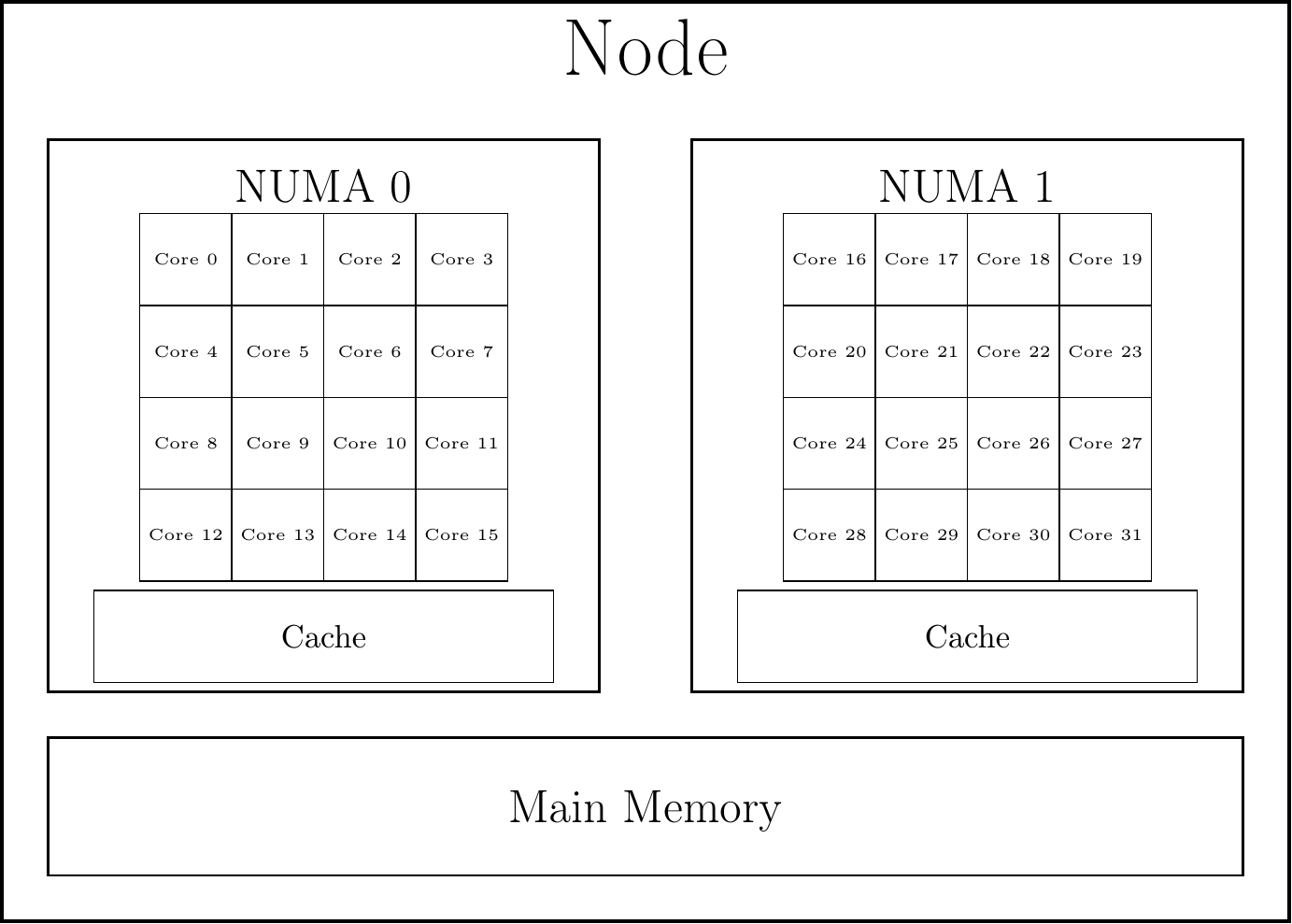}
    \caption{An example symmetric multiprocessing (SMP) node.}
    \label{fig:node}
\end{figure}

The remainder of this paper analyzes reductions to the cost of irregular communication through locality-aware neighbor collectives.  
Modern supercomputers contain a hierarchy of regions, with communication within a region being a different cost than between regions.  
For example, parallel architectures typically contain many nodes connected by a network, with each node containing many processes, as exemplified in Figure~\ref{fig:node}.
This example symmetric multiprocessing (SMP) node contains two non-uniform memory access (NUMA) regions, each with $16$ cores.  
All processes within a NUMA region share a level of cache, allowing intra-NUMA communication to be transferred through cache.  
Similarly, as all processes on the node share main memory, inter-NUMA communication within a node can be transferred through main memory.  
Finally, inter-node messages are injected into the network and transferred across the interconnect to the node of destination.  
As a result, computers achieve varying communication costs with regard to the locality of the messages.  
Locality-aware communication restructures point-to-point messages to reduce the most expensive messages in exchange for additional less costly communication.  

This paper introduces multiple novel strategies for adding locality-aware aggregation within neighborhood collectives, and presents significant associated performance improvements from replacing point-to-point communication within a widely used parallel codebase with locality-aware neighborhood collectives.  
The remainder of this paper is laid out as follows.  
Section~\ref{sec:bg} describes communication optimizations and neighborhood collectives in more detail and describes a number of related research works.  
Section~\ref{sec:impl} details the various neighborhood collective optimizations, and performance results associated with these implementations are presented in Section~\ref{sec:results}.  
Finally, conclusions and future directions are discussed in Section~\ref{sec:conc}.

\section{Background}~\label{sec:bg}
Each generation of supercomputer brings unique architectural design choices.  
In recent history, parallel systems have continuously increased potential compute power with additional complexity within each node.  
While older supercomputers such as the Blue Gene/L consisted of only a single dual-core chip per node~\cite{bgl}, Blue Gene/Q systems such as Sequoia, were comprised of symmetric multiprocessing (SMP) nodes with $16$ cores per node split across $2$ CPUs~\cite{bgq}.  
More recent systems, such as Summit, contain nodes with $2$ $22$-core CPUs~\cite{summit}, and emerging systems, such as Frontier, contain a single $64$-core chip per node, split into $4$ $16$-core NUMA regions~\cite{frontier}.  
The additional per-node complexity of each generation of parallel systems increases the variety in communication costs, with notable differences between intra-CPU, inter-CPU, and inter-node communication~\cite{PerfModelsP2P2018}.  
The performance differences between different regions of locality varies with systems, with inter-CPU but intra-node communication significantly more costly than inter-node on current and emerging systems~\cite{HeteroModeling2021}.

Parallel applications often fail to take full advantage of available compute power due to performance and scaling constraints associated with inter-process communication.  
Many simulations and numerical solvers are dominated by irregular communication, which requires each process to communicate varying amounts of data with a subset of other processes.  
Algebraic multigrid, for instance, relies on the performance of sparse matrix operations, such as the sparse matrix-matrix and sparse matrix-vector (SpMV) multiples.  
AMG first creates a hierarchy of increasingly dense matrices that approximate lower frequencies, with each successive matrix formed through a triple sparse matrix-matrix multiply.  
After the hierarchy is created, the solution is iteratively refined through numerous SpMVs on each level of the hierarchy.  
Sparse matrix operations require each process to receive data associated with every non-zero column held by the process.  
As a result, each process communicates varying amounts of data with a subset of other processes, as determined by the sparsity pattern of the given sparse matrix.  
As coarse levels within AMG are increasingly dense, communication requirements are often increased on levels near the middle of the hierarchy.  
Finally, at scale, the cost of these sparse matrix operations is dominated by the cost of irregular inter-process communication.  

\begin{exmp}\label{exmp:comm}
    Assume a system has multiple regions, each containing four processes, as displayed in Figure~\ref{fig:exm_comm}.  Each process within region 0 holds two unique values, represented as a circle and square.  The shaded regions of these objects correspond to the processes in region 1 to which each object must be sent.  For example, process $P0$ holds a circle shaded both red and green, and therefore must send this object to processes $P5$ and $P6$.  Furthermore, the square held by $P0$ is shaded blue, red, and teal, and therefore must be sent to processes $P4$, $P5$, and $P7$.  Throughout the remainder of this paper, the authors present multiple methods for communicating these values between regions $0$ and $1$.
    \begin{figure}[ht!]
    \centering
    \includegraphics{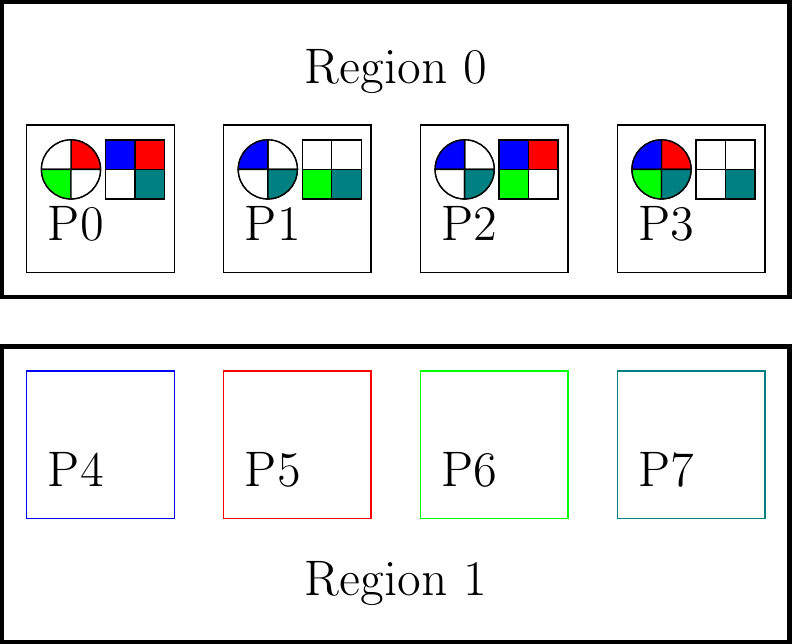}
    \caption{Example Communication}
    \label{fig:exm_comm}
\end{figure}
\end{exmp}
Example~\ref{exmp:comm} displays a simple irregular communication pattern.  
This paper shows the effects of locality-aware neighborhood collectives on communication throughout the iterative solve phase of AMG, as the associated SpMVs require a large range of communication patterns.  
Optimized neighbor collectives, however, are not limited to AMG and can be used to reduce the cost of irregular communication within other solvers and simulations.

Standard methods of irregular communication consist of gathering all data to be sent to a process before sending it directly as a single message.  
This approach fails to account for the locality of the sending and receiving processes.  
For instance, two processes within the same CPU are able to transfer data through cache, often at a significantly faster rate than data can be transported through the interconnect.  
Locality-aware methods, on the other hand, aggregate data within a region of locality to minimize the number and size of inter-region messages.  
The locality-aware neighborhood collectives presented in Section~\ref{sec:impl} utilize three-step aggregation, in which each process in a region communicates with a unique subset of other regions.  
The intra-region data is initially redistributed so that a single process per region holds all data to be sent to its unique subset of regions.  
Each process then sends a single message to each of its assigned regions.  
Finally, received data is redistributed within each region to transfer data to each final destination process.  
Note, there are many additional strategies for node-aware aggregation that could be utilized within neighborhood collectives.  
The authors focus on the three-step aggregation as this paper presents the effects of communication optimizations on sparse matrix-vector multiplication throughout AMG, where three-step aggregation has been shown to perform best~\cite{BienzLocAwareSpMV2019}.  
However, other simulations and solvers may be better optimized with additional locality-aware strategies.

While neighborhood collectives provide the necessary interface for optimizing irregular communication within MPI, they do require some overhead compared to standard point-to-point communication, namely with forming the neighborhood topology.  
Before a neighborhood collective, such as the \texttt{MPI\_Neighbor\_alltoallv}, can be executed, a neighborhood must first be formed.  For irregular communication, a neighborhood communicator can be formed at scale with the method \texttt{MPI\_Dist\_graph\_create\_adjacent}.  
This graph creation is passed data about each process that a given process sends to and receives from, and returns a directed neighborhood of processes with which each process communicates.  
There is synchronization overhead associated with this graph creation.  Only a single neighborhood is needed for each required communication pattern, however, such as each unique sparse matrix within a solver.  
Therefore, within iterative methods, the graph creation is amortized over subsequent iterations.

Persistent neighborhood collectives allow for further amortizations of setup costs across all iterations.  
Persistent MPI communication consists of initializing communication once, before starting and waiting on all communication at every iteration.  
Persistent neighborhood collectives first set up the collective with the \texttt{MPI\_Neighbor\_alltoallv\_init} method.  
Then, each iteration of communication consists of calls to \texttt{MPI\_Start} and \texttt{MPI\_Wait}, during which all communication is completed.  
Furthermore, the separate start and wait methods allow for an overlap of communication and computation, assuming the MPI implementation supports strong progress~\cite{10.1145/3416315.3416318}.  
This paper utilizes the persistent neighborhood API, allowing all locality-aware setup costs, such as load balancing while determining which intra-region process communicates with each region, to be incurred once within

\noindent\texttt{MPI\_Neighbor\_alltoallv\_init}.  
These overheads are then quickly offset by per-iteration reductions to communication costs.

\subsection{Related Work}
Before costly communication can be optimized, architectures and paths of communication must be accurately benchmarked and modeled for emerging systems to pinpoint the costs of the various messages.  
As emerging systems increase in complexity, performance models and benchmarks are adapted to fully capture the costs of the various paths of irregular communication.  
While the postal models accurately profile simple point-to-point communication~\cite{postal}, many extensions have been necessary to capture costs that dominate SMP architectures. 
For instance, the maxrate model greatly improves inter-node communication costs over the postal model by adding in measures for injection bandwidth limits~\cite{GroppMaxRate2016}.  
The maxrate model is further optimized through locality-awareness, modeling intra-CPU, inter-CPU, and inter-node messages separately~\cite{PerfModelsP2P2018}.  
While the maxrate model accurately captures costs of inter-node communication, intra-node communication models are further improved by adding constraints for all active processes, as bandwidth varies within a node based on the number of active processes~\cite{ThuneNodeP2PModels2023}.  
Finally, models for irregular communication, particularly for the large number of messages that occur within the coarse levels of AMG, are further improved by estimating queue search and network contention costs~\cite{PerfModelsP2P2018}.

Locality-aware communication has previously been explored extensively, both with point-to-point communication and throughout MPI collectives.  
Three-step aggregation, the focus of this paper, has shown to greatly improve instances of irregular communication in which many small messages are sent, such as in the solve phase of AMG~\cite{BienzLocAwareSpMV2019}.  
Similarly, two-step aggregation greatly reduces the costs associated with sending numerous larger messages such as within sparse matrix-matrix multiplies~\cite{BienzLocAwareAMG2020}, while ideal aggregation, which combines portions of messages ranging from two-step to three-step, optimizes the costs of medium-sized messages, such as within sparse matrix-multi-vector multiplies~\cite{LockhartLocAwareECG}.  
Similar aggregation techniques have shown large speedups within inter-GPU communication on heterogeneous architectures~\cite{HidayetogluLocAwareGPU, LockhartLocAwareGPU}.

Node-awareness is also a common technique for improving the performance of collective communication.  
Hierarchical communication consists of creating one or more master processes per node, and only performing steps of inter-node communication between these master processes~\cite{KaronisHierColl2000, GrahamHierCheetah2011, TraffHierAllgather2006, KandallaMultliLeaderHier2009}.  
Multi-lane approaches have further optimized inter-node communication within large collectives by having each process per node communicate a portion of the inter-node data~\cite{TraffMultilane2020}.  
Locality-aware collective algorithms reduce the cost of small collectives by minimizing the number of inter-node steps, having each process per node communicate with a separate node at each inter-node step~\cite{BienzLocAwareAllreduce2019, BienzLocAwareBruck2022}.

Topology-awareness, or optimizing algorithms for a given interconnect, is another common approach for minimizing collective communication costs.  
There are two categories of topology-aware algorithms, those which remap data to cores to minimize the number of hops messages are communicated~\cite{MirsadeghiTopoAwareRRCollectives2016, BhateleTorusColl2012, MaHierKNEMTopoAware2012}, and those that reformulate algorithms to minimize the number of steps for a given topology~\cite{PatarasukTreeAllreduce2007, OuyangMeshAllreduce}. 
While topology-aware algorithms greatly improve the performance of collective algorithms, they are specific to a given interconnect, which varies with emerging architectures.

There are a number of APIs for irregular communication that exist within the MPI 4 standard, and therefore implemented within all versions of MPI, including persistent and partitioned communication.  
Persistent communication reduces initialization costs by having an initialization so that all overhead is only incurred once~\cite{persistent}.  
All subsequent communications then communicate data without initialization overhead.  
Persistent communication exists for both point-to-point communication and collective operations.  
Partitioned communication extends the persistent point-to-point interface, allowing multiple threads or tasks to contribute data to a single message~\cite{partitioned, partitionedmpi4}.  
As a result, large messages that are partitioned across threads are sent in chunks rather than incurring a synchronization cost waiting for all threads to initialize corresponding communication.

\section{Persistent Neighborhood Collective Implementations}~\label{sec:impl}
Neighborhood collectives, such as the \texttt{MPI\_Neighbor\_alltoallv} provide the API for irregular communication optimizations within MPI.  
Furthermore, the persistent version of this method (released in MPI 4) allows for further optimizations because overhead, such as load balancing, is only incurred once and amortized over all successive iterations.  
Persistent neighbor collectives can wrap irregular communication throughout parallel applications, replacing point-to-point communication with a single initialization step, \\ \texttt{MPI\_Neighbor\_alltoallv\_init}, followed by \texttt{Start} and \texttt{Wait} to begin and complete each iteration of communication, respectively.  
All neighborhood collectives, regardless of persistence, do require an additional step of setup beyond point-to-point communication, as the topology communicator must be first be formed with a method such as \texttt{MPI\_Dist\_graph\_create\_adjacent}.

\subsection{Standard Neighborhood Collectives}~\label{sec:unopt}

The standard \texttt{MPI\_Neighbor\_alltoallv\_init} implementation consists of gathering all data to be sent to any process, and sending it directly, regardless of regions of sending and receiving processes, as displayed in algorithm~\ref{alg:standard}, in which \texttt{args} are the standard \texttt{MPI\_Neighbor\_alltoallv\_init} arguments.

\begin{algorithm2e}[ht!]
  \DontPrintSemicolon%
  \KwIn{$\texttt{args}$\tcc*{Standard method arguments}
        }

  \BlankLine%
    \tcp{Send count and processes, recv count and processes, in communicator}
    Get $n_{\texttt{send}}$, $p_{\texttt{send}}$, $n_{\texttt{recv}}$, $p_{\texttt{recv}}$ from communicator\;
    \BlankLine%

   \For{$i \gets 0$ \KwTo $n_{send}$}{ 
       MPI\_Isend\_init to $p_{\texttt{send}_{i}}$\tcc*{Send buffer, displacements, and size in args}\;
   }
   \For{$i \gets 0$ \KwTo $n_{recv}$}{ 
       MPI\_Irecv\_init from $p_{\texttt{recv}_{i}}$\tcc*{Recv buffer, displacements, and size in args}\;
   }
  
    \caption{Standard \texttt{standard\_init}}\label{alg:standard}
\end{algorithm2e}

Similarly, during each instance of communication, all messages are started at once, as shown in algorithm~\ref{alg:standard_start}.

\begin{algorithm2e}[ht!]
  \DontPrintSemicolon%
  \KwIn{$\texttt{args}$\tcc*{Standard method arguments}
        }

    \BlankLine%
    \tcp{Send count and processes, recv count and processes, in communicator}
    Get $n_{\texttt{send}}$, $p_{\texttt{send}}$, $n_{\texttt{recv}}$, $p_{\texttt{recv}}$ from communicator\;
    \BlankLine%
  
   \For{$i \gets 0$ \KwTo $n_{send}$}{ 
       Start send $i$\;
   }
   \For{$i \gets 0$ \KwTo $n_{recv}$}{ 
       Start recv $i$\;
   }
  
    \caption{Standard \texttt{standard\_start}}\label{alg:standard_start}
\end{algorithm2e}
The calling process then waits for all messages to complete, such as with \texttt{MPI\_Waitall}, as displayed in algorithm~\ref{alg:standard_wait}.

\begin{figure}[ht!]
    \centering
    \includegraphics{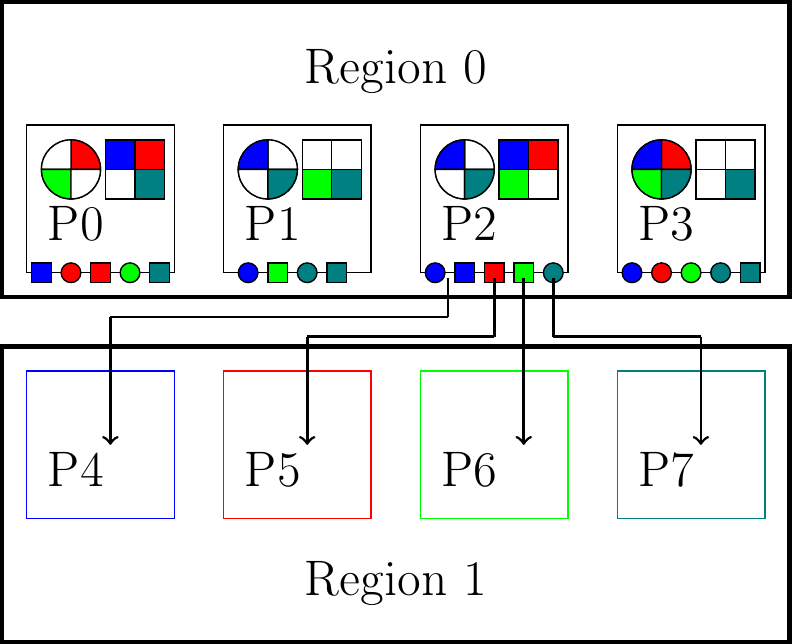}
    \caption{Standard Communication}
    \label{fig:exmp_standard_comm}
\end{figure}

\begin{algorithm2e}[ht!]
  \DontPrintSemicolon%
  \KwIn{$\texttt{args}$\tcc*{Standard method arguments}
        }

    \BlankLine%
    \tcp{Send count and recv count, in communicator}
    Get $n_{\texttt{send}}$, $n_{\texttt{recv}}$ from communicator\;
    \BlankLine%

   Wait for $n_{\texttt{send}}$ sends and $n_{\texttt{recv}}$ receives to complete
  
    \caption{Standard \texttt{standard\_wait}}\label{alg:standard_wait}
\end{algorithm2e}
Standard implementations directly wrap point-to-point messages within a single API.  
They fail to optimize the communication, however, by e.g. minimizing expensive communication between non-local regions.  
For instance, standard neighborhood collective communication of Example~\ref{exmp:comm} consists of each process in region 0 communicating with all processes in region 1, as displayed in Figure~\ref{fig:exmp_standard_comm}.  
This figure displays all messages originating on process $P2$.  
This procedure gathers both values represented by the circle and square, and sends them in a single messages to process $P4$, as both shapes on $P2$ have shaded blue regions, indicating $P4$ requires both values.  
The value represented by the square is then additionally sent multiple times, once to $P5$ and once to $P6$.  
Finally, the value represented by the circle is also sent to $P7$.  
In total, this example requires 15 messages to be sent from region 0 to region 1, and all data values with multiple indicating colors are sent in multiple inter-region messages.

\subsection{Aggregating Messages}~\label{sec:part}
The \texttt{MPI\_Neighbor\_alltoallv\_init} method provides the necessary information for locality-aware optimizations, such as aggregation of data within local regions, to be performed within MPI libraries.  
The method parameters include information on all processes with which to send or receive data, and may optionally include the amount of data to be sent to each.  
This is sufficient information for all processes within a region to determine the regions to which they send to and receive from.  
If weights are passed to the method, the regions are also able to determine inter-region data sizes.  
Methods of aggregation, such as locality-aware strategies, partition the communication across all processes per region so that each sends a minimal portion of messages for small data sizes, or an equal portion of data when sizes are large.

Aggregation within the persistent neighborhood collectives is shown in algorithm~\ref{alg:aggregated_init}. 
The black text, excluding red text, describes aggregation within the standard \texttt{MPI\_Neighbor\_alltoallv\_init} API.

\begin{algorithm2e}[ht!]
  \DontPrintSemicolon%
  \KwIn{$\texttt{args}$\tcc*{Standard method arguments}
        \textcolor{red}{$\texttt{send\_idx}$\tcc*{Unique send indices}
        $\texttt{recv\_idx}$\tcc*{Unique recv indices}}
        }

    \BlankLine%
    \tcp{Send count and processes, recv count and processes, in communicator}
    Get $n_{\texttt{send}}$, $p_{\texttt{send}}$, $n_{\texttt{recv}}$, $p_{\texttt{recv}}$ from communicator\;
    \BlankLine%

    \tcp{Form aggregated communication args}
    \tcp{$l$ : fully local communication, $s$ : initial local redistribution}
    \tcp{$g$ : global, inter-region communication, $r$ : final redistribution of received data}
    \texttt{setup\_aggregation($l$, $s$, $g$, $r$\textcolor{red}{, \texttt{send\_idx}, \texttt{recv\_idx}})}\;
    
    \BlankLine%

    \texttt{standard\_init}($l$)\;
    \texttt{standard\_init}($s$)\;
    \texttt{standard\_init}($g$)\;
    \texttt{standard\_init}($r$)\;
  
    \caption{Aggregated \texttt{aggregated\_init}}\label{alg:aggregated_init}
\end{algorithm2e}

The method \texttt{setup\_aggregation} creates the path of aggregation, assigning a portion of the inter-region communication to each process within a region.  
All examples and results presented in this paper use three-step aggregation, but this could be replaced by any aggregation strategy.    

The aggregated communication is split into four separate steps : 
\begin{itemize}
    \item $\ell$ : fully local communication, with source and destination process located within the same region
    \item $s$ : initial redistribution of data within a region
    \item $g$ : inter-region communication
    \item $r$ : final redistribution of received data within a region
\end{itemize}
Persistent communication is initialized for each of these four steps.

During each instance of communication, all fully local $\ell$ and inter-region $g$ communication is started within the method \texttt{start}, as described in algorithm~\ref{alg:agg_start}.

\begin{algorithm2e}[ht!]
  \DontPrintSemicolon%
  \KwIn{$l_{\texttt{args}}$\tcc*{fully local communication arguments}
  $s_{\texttt{args}}$\tcc*{initial local redistribution arguments}
  $g_{\texttt{args}}$\tcc*{global, inter-region communication arguments}
  $r_{\texttt{args}}$\tcc*{final local redistribution arguments}
        }

  \tcp{Start fully local communication}
  \texttt{standard\_start}($l_{\texttt{args}}$)\;

  \tcp{Start and complete initial redistribution}
  \texttt{standard\_start}($s_{\texttt{args}}$)\;
  \texttt{standard\_wait}($s_{\texttt{args}}$)\;

  \tcp{Start inter-region communication}
  \texttt{standard\_start}($g_{\texttt{args}}$)\;

    \caption{Aggregated \texttt{aggregated\_start}}\label{alg:agg_start}
\end{algorithm2e}
The initial redistribution of data within the region must be fully completed before inter-region communication can begin.  
Therefore, this method consists of both starting and completing the initial redistribution $s$, before starting the inter-region communication $g$.

Finally, each instance of communication is completed within the \texttt{wait} method, described in algorithm~\ref{alg:agg_wait}.

\begin{algorithm2e}[ht!]
  \DontPrintSemicolon%
  \KwIn{$l_{\texttt{args}}$\tcc*{fully local communication arguments}
  $s_{\texttt{args}}$\tcc*{initial local redistribution arguments}
  $g_{\texttt{args}}$\tcc*{global, inter-region communication arguments}
  $r_{\texttt{args}}$\tcc*{final local redistribution arguments}
        }

  \tcp{Complete fully local communication}
  \texttt{standard\_wait}($l_{\texttt{args}}$)\;

  \tcp{Complete inter-region communication}
  \texttt{standard\_wait}($g_{\texttt{args}}$)\;

  \tcp{Start and complete final redistribution}
  \texttt{standard\_start}($r_{\texttt{args}}$)\;
  \texttt{standard\_wait}{$r_{\texttt{args}}$}\;

    \caption{Aggregated \texttt{aggregated\_wait}}\label{alg:agg_wait}
\end{algorithm2e}
The inter-region communication $g$ must complete before the final intra-region redistribution of data can be performed.  
Therefore, this method consists of completing the inter-region step $g$, before both starting and completing the final intra-region redistribution $r$.  

\begin{figure}[ht!]
    \centering
    \includegraphics{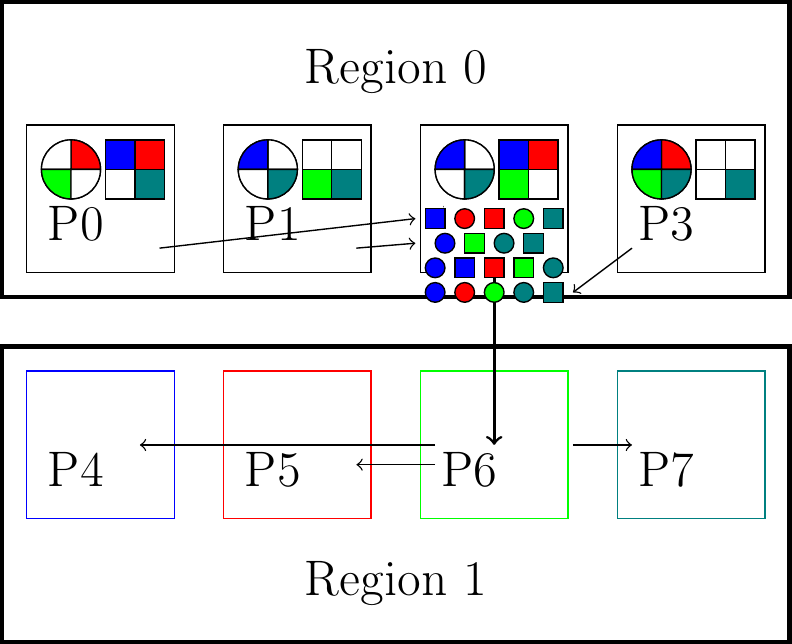}
    \caption{Partially Aggregated Communication}
    \label{fig:exmp_partial_comm}
\end{figure}

This approach greatly reduces the number of inter-region messages.  
For instance, the inter-region communication required within Example~\ref{exmp:comm} is performed in three steps, as shown in Figure~\ref{fig:exmp_partial_comm}.
Initially, all inter-region messages are redistributed locally so that each process holds all data to be sent to a unique subset of regions.  
For example, in Figure~\ref{fig:exmp_partial_comm}, all data to be communicated to region 1 is first sent locally to process $P2$.  
Then, process $P2$ sends a single inter-region message to $P6$.  
Finally, process $P6$ redistributes the received values locally.

\subsection{Extensions for Duplicate Data}~\label{sec:full}
While the persistent neighborhood collective API provides sufficient information to aggregate messages within each region, it fails to include necessary information for removal of duplicate values.  
For instance, Example~\ref{exmp:comm} displays a scenario in which many processes in region 0 are sending the same value to multiple destination processes within region 1.  
In all previous approaches, these values are communicated between the region duplicate times, once per destination.  
As the current API does not provide information on the indices being communicated, there is no way to remove these duplicates. 

A small extension to the API, requiring unique indices associated with each data value to be communicated, would allow for minimization of inter-region message sizes.  
This extension to \texttt{MPI\_Neighbor\_alltoallv\_init} is displayed as red text throughout Algorithm~\ref{alg:aggregated_init}.  
The extra information can be used while setting up aggregation to remove duplicate values from inter-region communication.  
Note, while all aggregation results in this paper are based on three-step node-aware strategies, any aggregation technique could use this information to minimize inter-region data sizes.

\begin{figure}[ht!]
    \centering
    \includegraphics{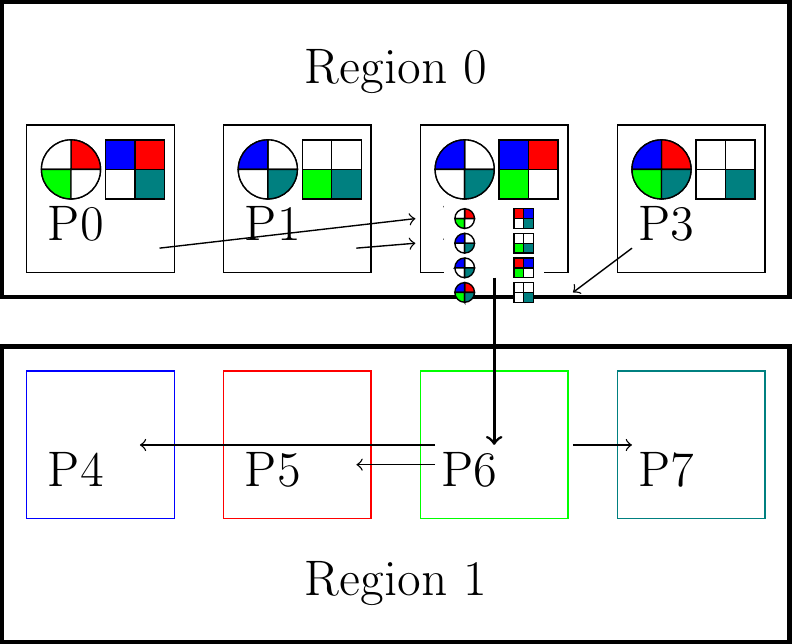}
    \caption{Fully Aggregated Communication}
    \label{fig:exmp_full_comm}
\end{figure}

Figure~\ref{fig:exmp_full_comm} displays the fully aggregated approach for communicating the values in Example~\ref{exmp:comm}.  
As discussed, each value from each source process now has only one copy sent both within the two regions for aggregation and between the two regions.

\section{Experimental results}~\label{sec:results}
The performance of the neighborhood collective implementations presented in Section~\ref{sec:impl} are analyzed throughout the sparse matrix-vector multiplies of the solve phase of Hypre's BoomerAMG.  
All experiments are performed on a $7$-point rotated anisotropic diffusion system, with rotated of $45$ degrees and anisotropy of $0.001$.  

All experiments were run on the CPU cores of Lassen, a Power9 system at Lawrence Livermore National Laboratory, using the system install of Spectrum MPI.  
Each node of Lassen contains two 22-core CPUs.  
While intra-CPU communication outperforms inter-node, inter-CPU communication within a node requires over twice the cost of inter-node for large messages~\cite{HeteroModeling2021}.  
Therefore, all presented results use only $16$ cores per node on a single CPU to avoid inter-CPU expenses.  

In an effort to achieve performance reproducibility, each performance result presented in this section acquires the time required to perform $1000$ calls to \texttt{MPI\_Start} and \texttt{MPI\_Wait}, and then finds the average cost of a single instance of those $1000$ steps of communication.  
Each test is run three separate times and the minimum of the three resulting averages is taken, reducing the impact of nearby jobs.

Throughout the presented results, the following four communication protocols are analyzed : 
\begin{itemize}
    \item Standard Hypre : persistent point-to-point communication as implemented in release 2.28 of Hypre
    \item Unoptimized neighborhood collectives : standard communication within a persistent neighborhood collective, as described in Section~\ref{sec:unopt}.
    \item Partially optimized neighborhood collectives : locality-aware aggregation within a persistent neighborhood collective, as described in Section~\ref{sec:part}
    \item Fully optimized neighborhood collectives : locality-aware aggregation plus the removal of duplicate values, as described in Section~\ref{sec:full}
\end{itemize}
All neighborhood collective implementations are implemented within a lightweight open source library, MPI Advance, that is then linked with Hypre.  
Implementations within MPI Advance then call necessary instances of point-to-point communication using the system install of MPI.

\begin{figure}[ht!]
    \centering
    \includegraphics[width=\linewidth]{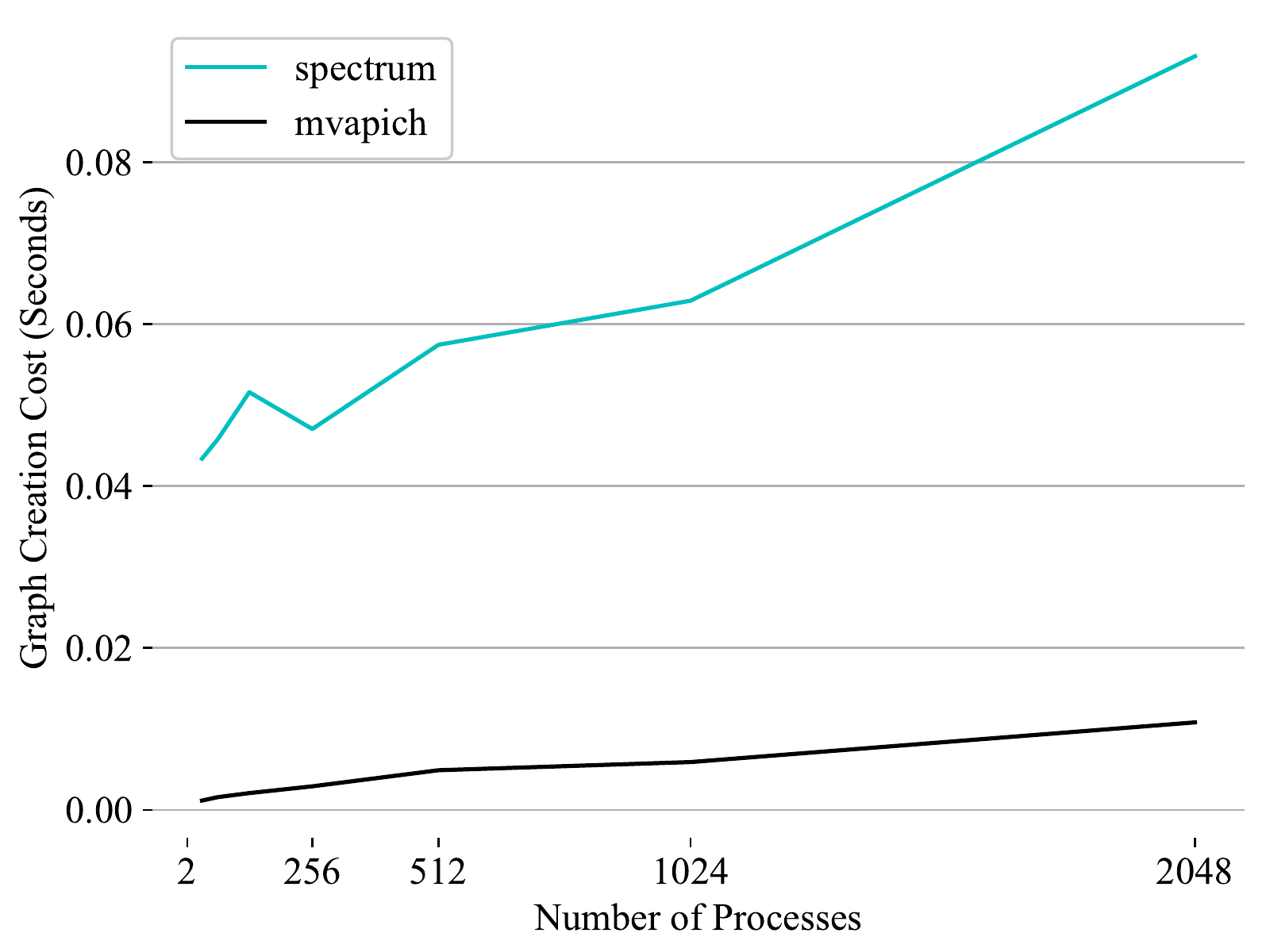}
    \caption{Cost of calling \texttt{MPI\_Dist\_graph\_create\_adjacent} once per level of the AMG hierarchy at a variety of process counts.  This problem is strongly scaled, with each rotated anisotropic diffusion system containing $524\,288$ rows.}
    \label{fig:dist_graph}
\end{figure}

There is an overhead to using neighborhood collectives over point-to-point communication, namely in creating the topology communicator.  
Neighborhood collectives require creating the topology communicator only once, amortizing this cost over all iterations of communication. 
For irregular communication, this communicator can be formed with the method \texttt{MPI\_Dist\_graph\_create\_adjacent}. 
The cost of this method was evaluated for two MPI implementations, Spectrum MPI and MVAPICH over a range of process counts in Figure~\ref{fig:dist_graph}. 
As shown in Figure~\ref{fig:dist_graph}, the method can be called with minimal overhead, but the choice of MPI implementation is important. 
For the problem tested, MVAPICH performs the method 8.6x as fast as Spectrum MPI at the scale of $2048$ cores.
The cost with MVAPICH also demonstrates improved strong scaling.

\begin{figure}[ht!]
    \centering
    \includegraphics[width=\linewidth]{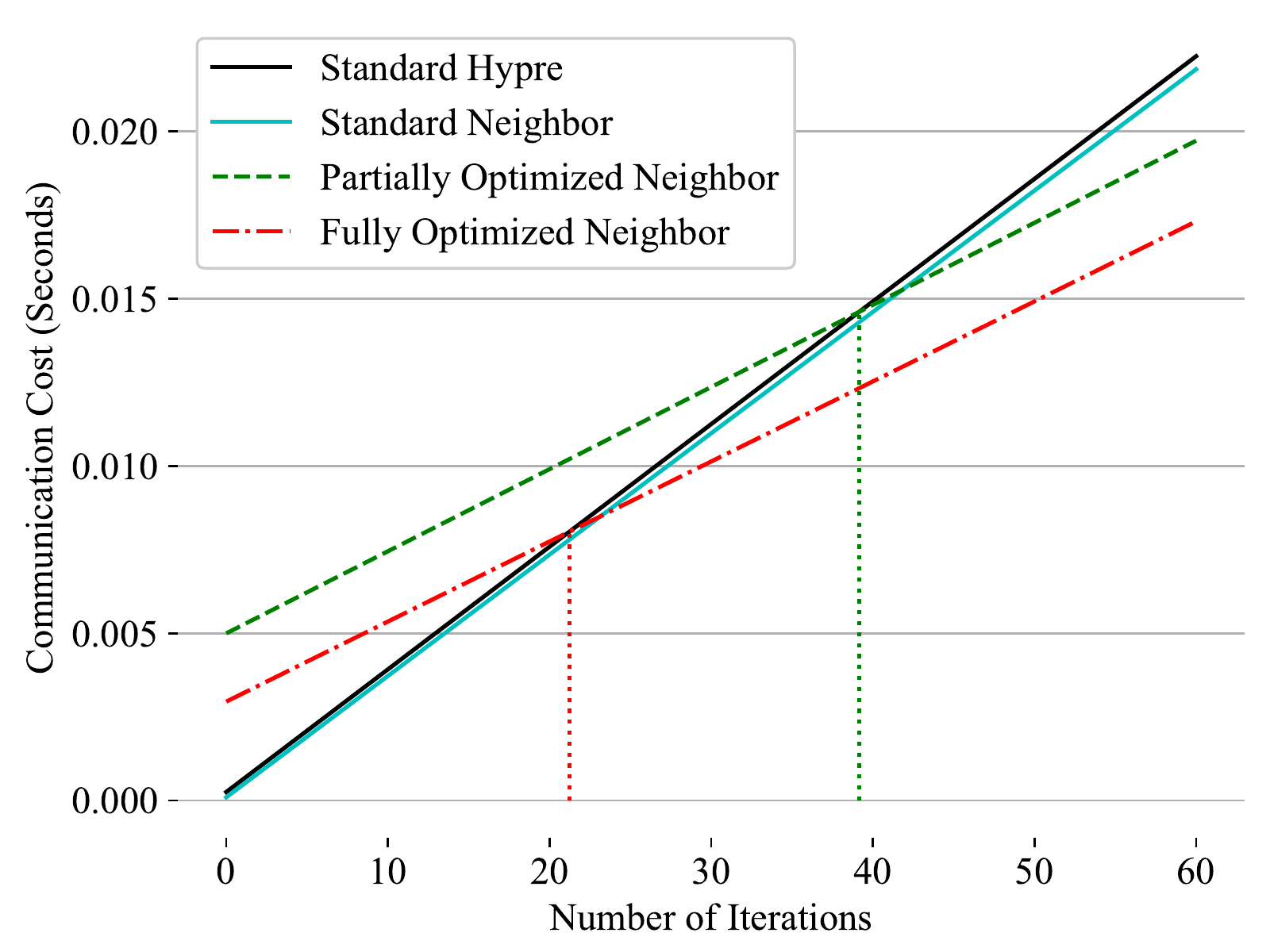}
    \caption{Cost of communication for a range of iteration counts. At each iteration count, cost is  calling \texttt{MPI\_Neighbor\_alltoallv\_init} once per level of the AMG hierarchy, plus calling \texttt{MPI\_Start} and \texttt{MPI\_Wait} once per AMG level per iteration.  This problem is a rotated anisotropic diffusion system containing $524\,288$ rows run on $2048$ cores}
    \label{fig:init}
\end{figure}

Persistent neighborhood collectives incur all setup costs only once during the initialization method.  
As a result, costly setup of optimizations, such as load balancing inter-region communication across all processes within a region, are amortized over all iterations of communication.  
Figure~\ref{fig:init} displays the costs associated with initializing each of the neighborhood collectives for a rotated anisotropic diffusion system containing $524\,288$ rows run on $2048$ cores.
The figure shows the cost of communication for a number of iterations added to the initialization cost across a range of iteration counts. 
Intersections, denoted by dotted vertical lines, indicate the number of iterations at which the higher initialization cost is outweighed by a lower per-iteration communication cost.
The crossover points found are 40 iterations for the partially optimized implementation, and 22 iterations for the fully optimized implementation.
There is minimal cost associated with the standard neighborhood implementation, as this method simply wraps point-to-point communication.
The partially optimized implementation demonstrates a higher initialization cost than the fully optimized implementation because the former simply wraps the latter.
The partially optimized initialization time could be further reduced by implementing it directly.
The overheads associated with aggregated communication techniques are due to forming the aggregated path of communication and load balancing.  
Note, as this cost is only incurred once per communication pattern, more significant initialization overheads are acceptable for higher iteration counts.  
For communication with fewer iterations, however, simpler aggregation techniques will be necessary to reduce initialization overheads.

\subsection{Per-Level Analysis}~\label{sec:per-level}
Algebraic multigrid requires sparse matrix operations to be performed across a variety of levels, with each level decreasing in dimension but often increasing in density.  
As a result, communication dominates coarse levels near the middle of the hierarchy.  
Locality-aware neighbor collectives reduce the inter-region message count and sizes in exchange for additional intra-region communication.  
This section analyzes the impact of locality-aware neighborhood collectives on each level of a rotated anisotropic diffusion hierarchy.  
The fine-level system contains $524\,288$ split across $2048$ cores.

\begin{figure}
    \centering
    \includegraphics[width=\linewidth]{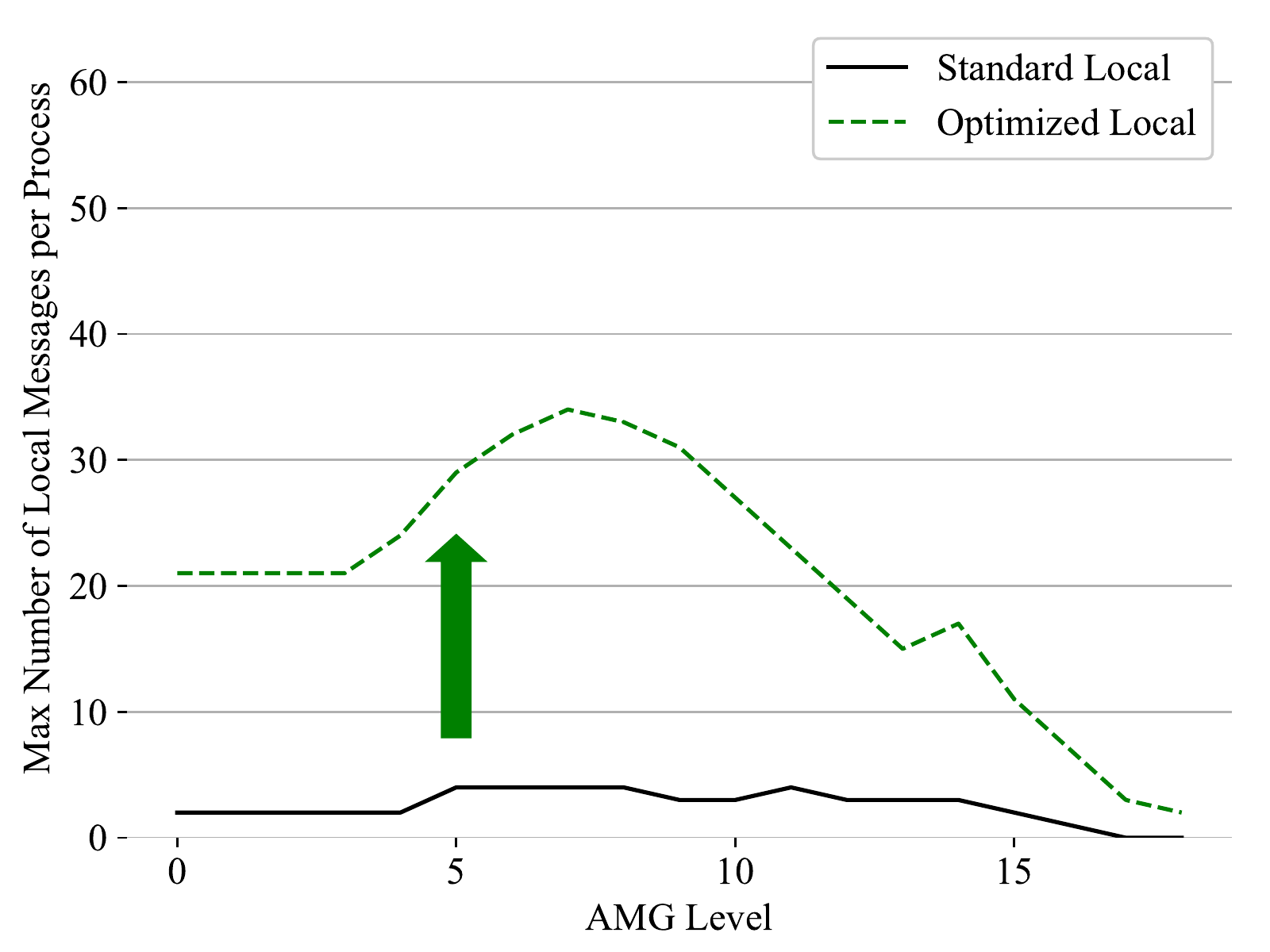}
    \caption{Per-level \textbf{intra-region message counts} when performing a SpMV on each level of the hierarchy.}
    \label{fig:msgs_local}
\end{figure}
Figure~\ref{fig:msgs_local} displays the maximum number of intra-region messages sent by any process on each level of the hierarchy.  
Locality-aware neighbor collectives greatly increase the intra-region communication requirements, as both initial and received data is redistributed among processes within each region.

\begin{figure}
    \centering
    \includegraphics[width=\linewidth]{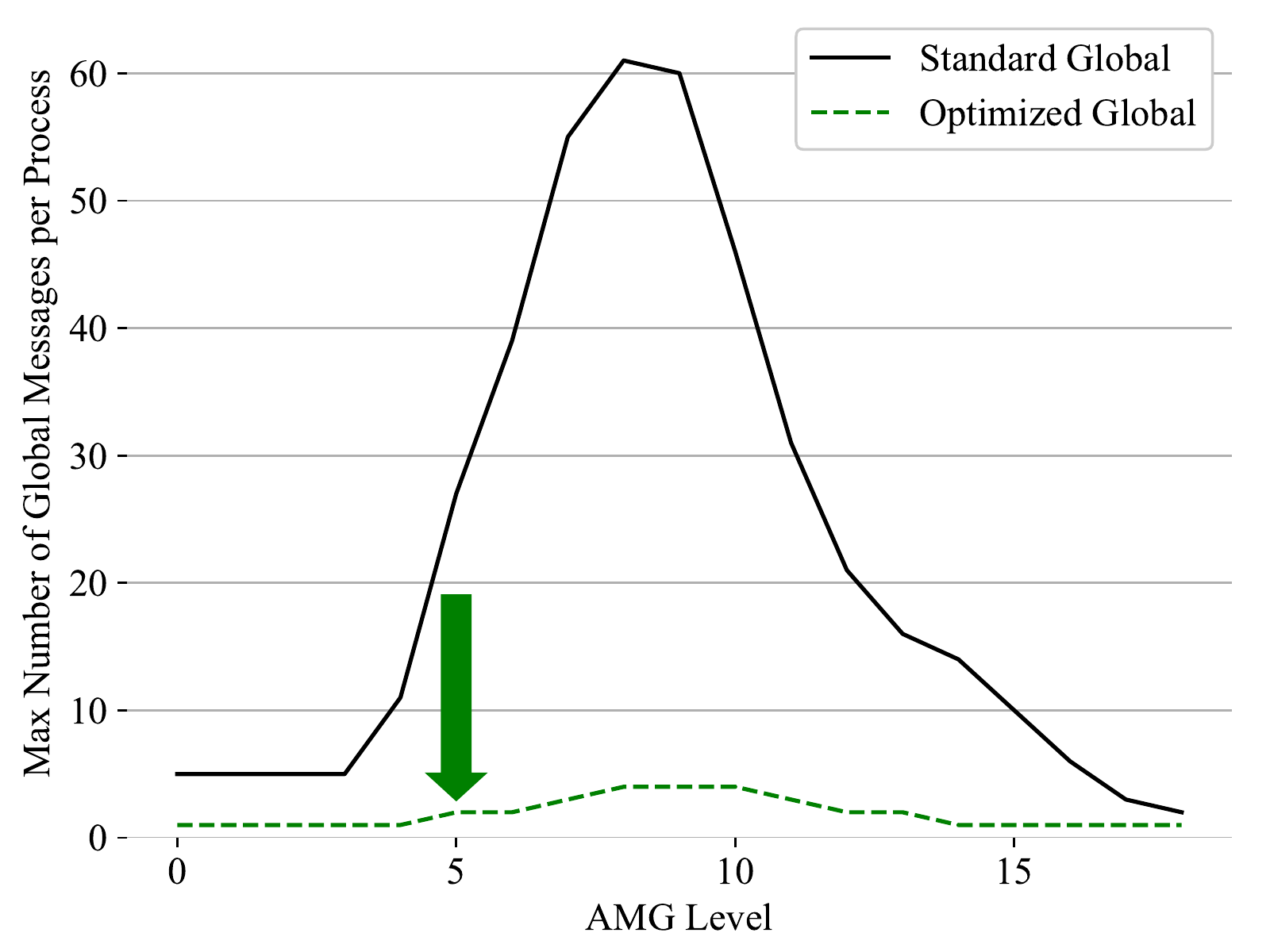}
    \caption{Per-level \textbf{inter-region message counts} when performing a SpMV on each level of the hierarchy.}
    \label{fig:msgs_global}
\end{figure}
Figure~\ref{fig:msgs_global} displays the maximum number of inter-region messages sent by any process on each level of the hierarchy.  
While locality-aware aggregation greatly increased intra-region message counts, it results in a similar decrease in the more costly inter-region communication.

\begin{figure}
    \centering
    \includegraphics[width=\linewidth]{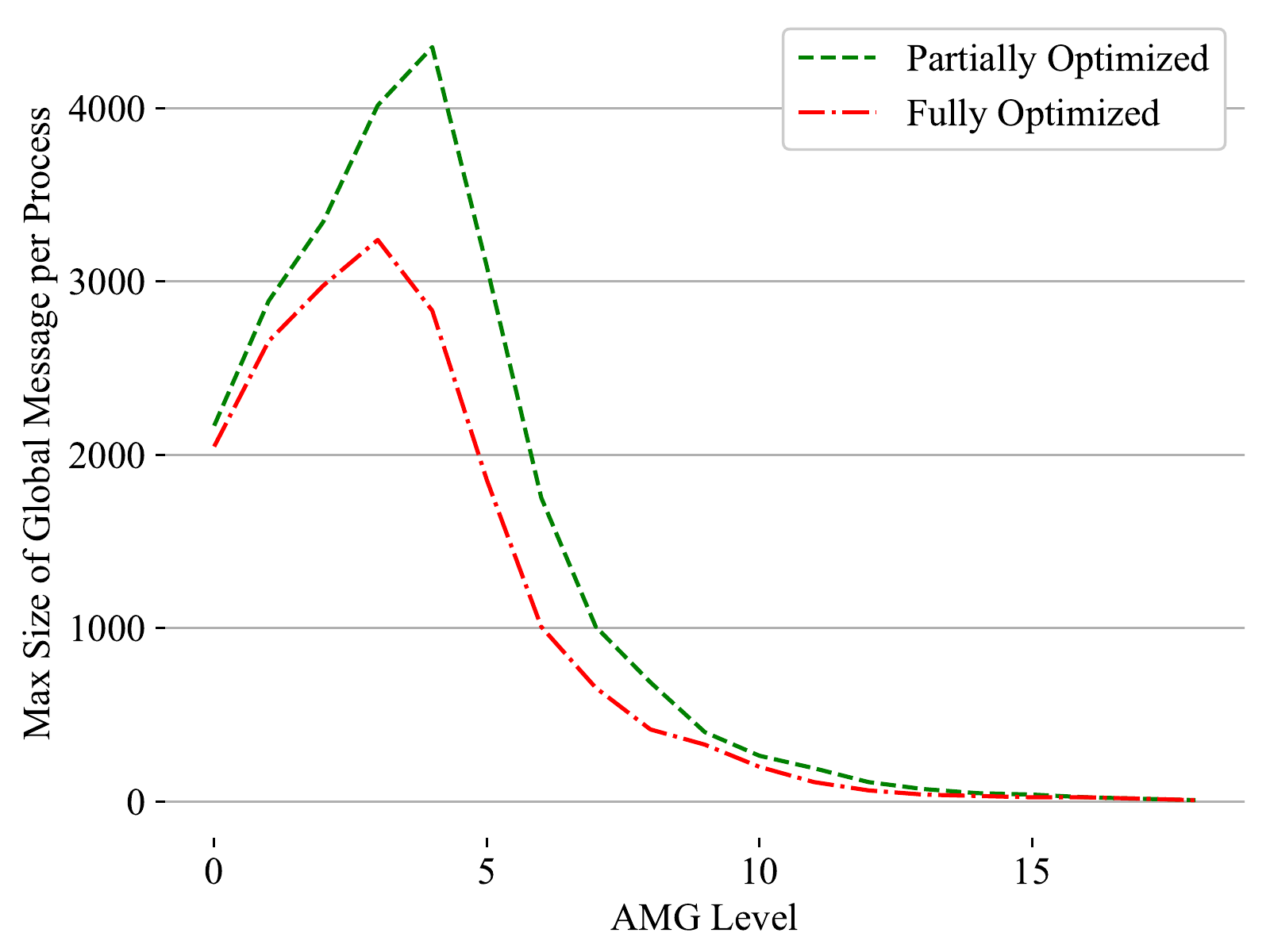}
    \caption{Per-level \textbf{inter-region message sizes} when performing a SpMV on each level of the hierarchy.}
    \label{fig:msg_size_global}
\end{figure}
Standard and partially optimized communication techniques result in data values being communicated multiple times between regions.  
Fully optimized neighbor collectives eliminate values from being communicated more than once between any set of regions.  
Figure~\ref{fig:msg_size_global} displays the per-level message sizes for the partially versus fully optimized neighbor collectives.
As shown, locality-aware deduplication results in up to a 35\% reduction of the maximum size of global messages per process for level 4 of the AMG hierarchy for the same $524\,288$ row problem tested on $2048$ processes.

\begin{figure}[ht!]
    \centering
    \includegraphics[width=\linewidth]{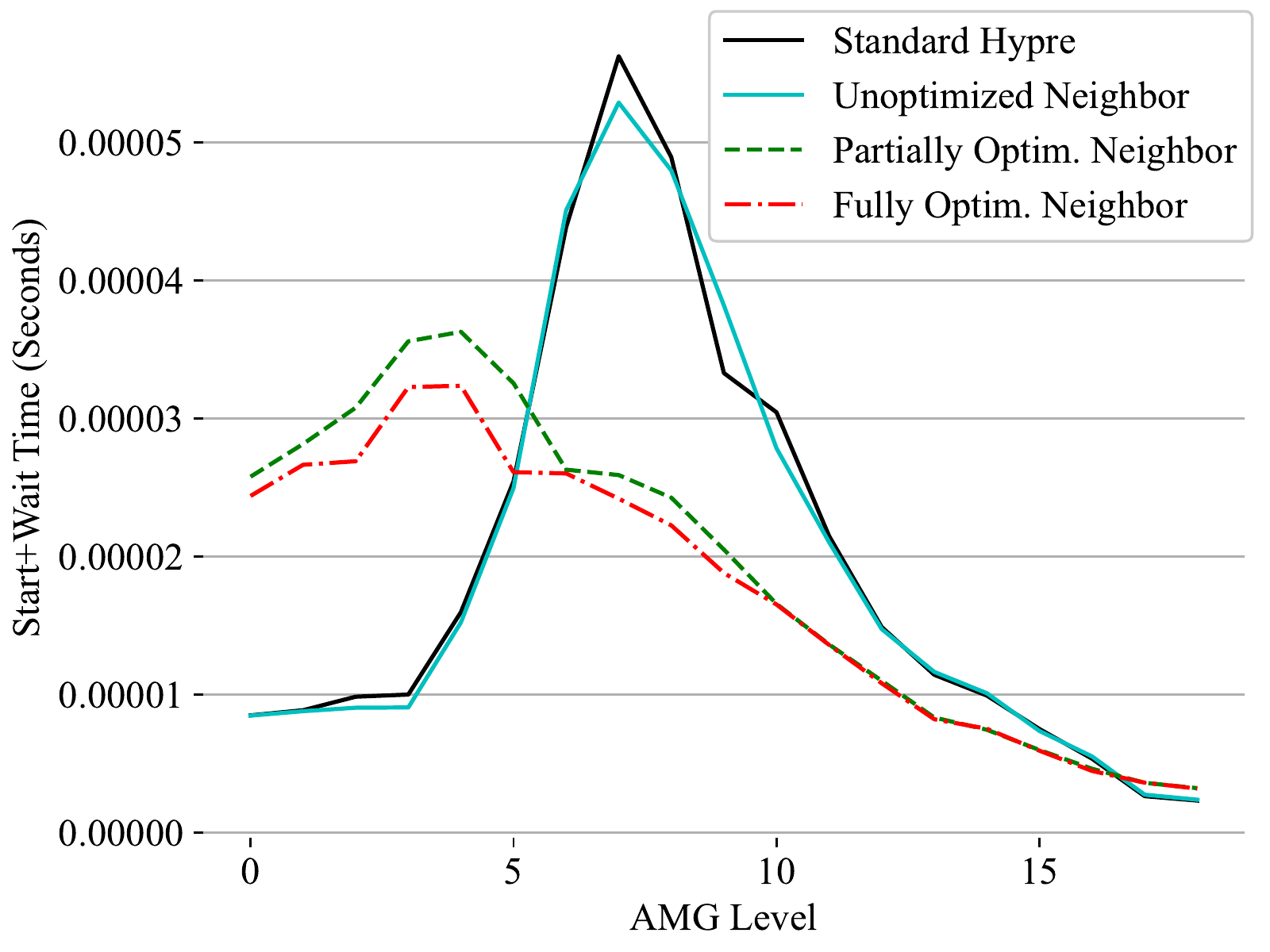}
    \caption{The cost of communicating data during a SpMV on each level of a rotated anisotropic diffusion hierarchy.  The fine-level system contains $524\,288$ rows split across $2048$ cores.}
    \label{fig:per_lvl}
\end{figure}

Figure~\ref{fig:per_lvl} displays the cost of communication within a sparse matrix-vector multiply on each level of the AMG hierarchy.  
Fine levels incur minimal communication overheads, as they are relatively sparse.  
Overheads incurred during local redistributions of data increases the cost of locality-aware neighborhood collectives over standard point-to-point communication on these levels.  
However, as per-level costs increase on the coarse levels, locality-aware aggregation techniques pay off, with optimized neighborhood collectives greatly outperforming standard communication near the middle of the hierarchy.  
Finally, there are additional benefits to removing values from being communicated multiple times between a single set of regions.  
Note, the coarsest levels are small enough in dimension that few processes participate in communication, resulting in minimal differences between communication strategies.

\subsection{Scaling Analysis}
The cost of communication, and performance of the various neighborhood collective implementations, varies with problem scale.  
This section analyzes the cost of communicating within a SpMV on every level of the AMG hierarchy at various scales.  
At each scale, the timing is a sum of the times required to perform SpMV communication on each level of the hierarchy at the given scale.  
The partially and fully optimized neighborhood results use the standard communication strategy on finer levels when it outperforms the locality-aware optimizations, summing up the least expensive of standard communication and the given optimized neighbor collective at each step.  
This demonstrates the maximum possible improvement over standard communication techniques.  
However, to achieve this performance, a selection strategy, such as a simple performance model, is needed to dynamically choose the optimal neighborhood collective implementation for a given communication pattern.

\begin{figure}
    \centering
    \includegraphics[width=0.5\textwidth]{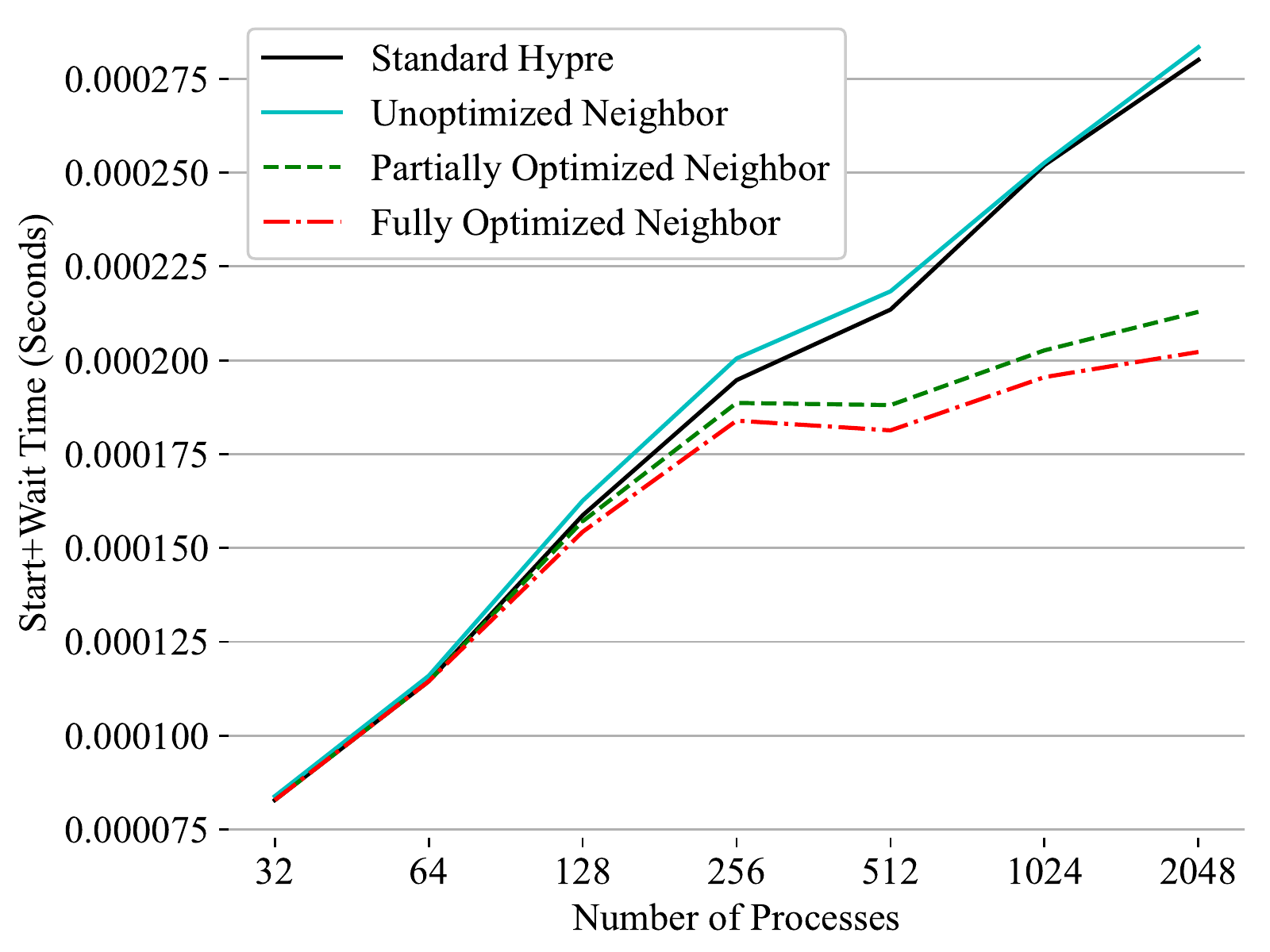}
    \caption{The cost of communicating within a SpMV on every level of the AMG hierarchy for a strongly scaled rotated anisotropic diffusion system with $524\,288$ rows.}
    \label{fig:strong}
\end{figure}
Figure~\ref{fig:strong} presents a strong scaling study of communication costs for a rotated anisotropic diffusion system with $524\,288$ rows split across process counts ranging from $64$ to $2048$. 

The unoptimized neighborhood collective performs similarly to the standard point-to-point communication within Hypre, with both strategies communicating equivalent message counts and sizes.  
The partially optimized neighbor collective significantly improves the scalability of this communication, achieving a speedup of 1.32x over standard communication at $2048$ processes.  
The fully optimized neighbor collectives achieve an additional 0.07x speedup by reducing the size of inter-region communication.  
As the problem is strongly scaled, the impact of the locality-aware neighborhood collectives increases, indicating that the optimized neighbor collective have increasingly large impacts and message counts increase, and per-message sizes decrease.

\begin{figure}
    \centering
    \includegraphics[width=\linewidth]{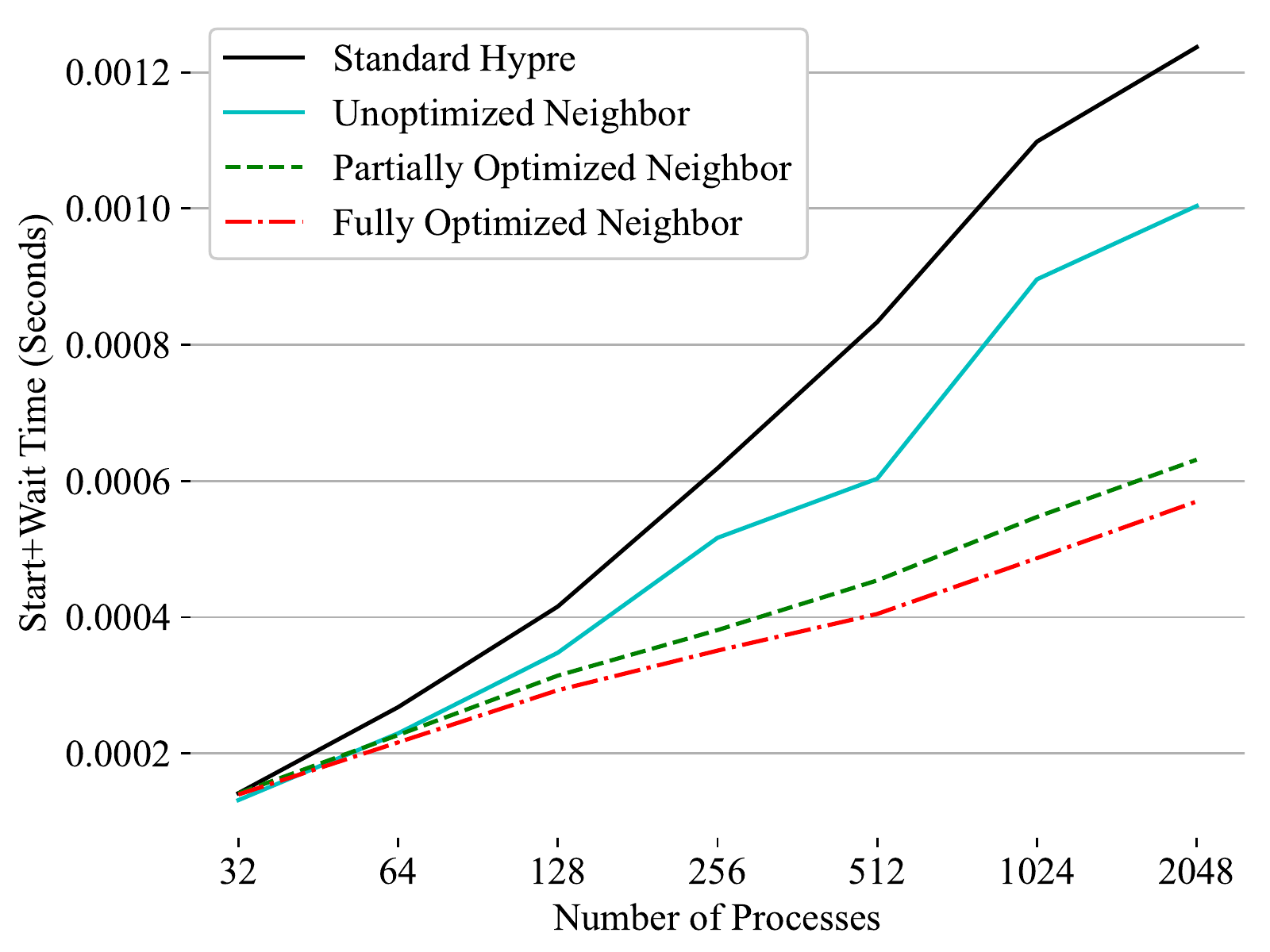}
    \label{fig:weak}
    \caption{The cost of communicating within a SpMV on every level of the AMG hierarchy for a weakly scaled rotated anisotropic diffusion system with $524\,288$ rows.}
\end{figure}
Figure~\ref{fig:weak_scale} presents weak scaling results for the communication on each level of a rotated anisotropic diffusion hierarchy with $524\,288$ rows, scaling from $64$ to $2048$ cores. 
The weak scaling study shows that the impact of locality-aware aggregation increases with process count.  
As a larger number of processes are performing communication, there is an additional benefit to reducing duplicate messages between processes.
For the weakly scaled problem at $2048$ cores, locality-aware aggregation results in a speedup of 1.96x, while reducing duplicate messages provides an additional 0.21x speedup.

\section{Conclusions and Future Directions}~\label{sec:conc}
Persistent neighborhood collectives provide the interface for locality-aware optimizations to be efficiently implemented within MPI.  
Standard point-to-point communication can be efficiently replaced with neighbor collectives, incurring only an additional overhead associated with graph creation.  
However, while forming the topology communicator has a significant cost at scale, this cost is only incurred once and then amortized over all iterations of communication.  
Similarly, initialization of locality-aware aggregation techniques can incur large overheads.  
However, the persistent neighborhood collective only requires this initialization to occur once per communication pattern, before also being amortized over all iterations of communication.

Locality-aware neighbor collectives, implemented in MPI Advance, significantly improve the performance of irregular communication throughout the coarse levels of Hypre, in which communication requirements are the largest.  
Furthermore, the optimized neighbor collectives improve both weak and strong scalability of the solver when the appropriate communication strategy is selected at each level of the hierarchy.  
Eliminating values from being communicated multiple times between a single set of regions further increases these improvements.  
Finally, as the neighborhood collective implementations exist within MPI Advance, they are accessible to all applications that are limited by irregular communication, requiring the application only to replace point-to-point communication with neighborhood collective and link with MPI Advance.

While locality-aware neighbor collectives have the potential to greatly improve performance, they also are capable of greatly increasing communication costs, particularly for patterns with fewer communication requirements.  
As a result, a simple performance measure is needed within the neighborhood collective to dynamically select the optimal communication strategy.  
Furthermore, there are many existing aggregation techniques for locality-aware communication not discussed in this paper.  
Additional aggregation strategies should be added into MPI Advance, allowing for the dynamic selection not only of locality-aware aggregation, but also of the optimal type of aggregation.  
Finally, other optimizations should also be added within the implementations of neighborhood collectives.  
For instance, large messages have been optimized separately with both locality-aware methods and partitioned communication~\cite{partitionedmpi4}.  
The combination of these optimizations, partitioning locality-aware messages, can have an even large impact on communication requirements.

Currently, neighborhood collective implementations optimize only inter-CPU communication.  
State-of-the-art computers such as Lassen, however, consist of heterogeneous nodes with multiple GPUs per node.  
Many applications, such as Hypre, achieve full performance through acceleration on these GPUs, relying on inter-GPU communication.  
Neighborhood collective strategies can be extended to optimize inter-GPU communication, not only dynamically selecting the optimal locality-aware strategy, but also determining whether to communication data directly between GPUs, to first copy to CPUs, or to copy a portion of the data to each available CPU core, allowing each to communicate with a smaller subset of regions.

\section{Acknowledgements}
This work was performed with partial support from the National Science
Foundation under Grant No. CCF-2151022 and the U.S. Department of Energy's National Nuclear Security Administration (NNSA) under the Predictive Science Academic Alliance Program (PSAAP-III), Award DE-NA0003966.

Any opinions, findings, and conclusions or recommendations expressed in this material are those of the authors and do not necessarily reflect the views of the National Science Foundation and the U.S. Department of Energy's National Nuclear Security Administration.

\bibliographystyle{siamplain}
\bibliography{references}
\end{document}